\newcommand{\CLASSINPUTinnersidemargin}{0.65in}
\newcommand{\CLASSINPUTbottomtextmargin}{0.8in}
\newcommand{\CLASSINPUTtoptextmargin}{0.8in}

\documentclass[conference]{IEEEtran}
\makeatletter
\def\ps@headings{%
\def\@oddhead{\mbox{}\scriptsize\rightmark \hfil \thepage}%
\def\@evenhead{\scriptsize\thepage \hfil \leftmark\mbox{}}%
\def\@oddfoot{}%
\def\@evenfoot{}}
\makeatother
\usepackage{cite}

\ifCLASSINFOpdf
  \usepackage[pdftex]{graphicx}
\else
  \usepackage[dvips]{graphicx}
\fi
\usepackage[cmex10]{amsmath}
\usepackage{mathptmx}
\usepackage{bm}

\usepackage[tight,footnotesize]{subfigure}
\usepackage{url}

\begin{document}
%
\title{Characterizing Phishing Threats with Natural Language Processing}

\author{\IEEEauthorblockN{Michael C. Kotson}
\IEEEauthorblockA{MIT Lincoln Laboratory\\
Lexington, MA\\
Email: Michael.Kotson@ll.mit.edu}
\and
\IEEEauthorblockN{Alexia Schulz}
\IEEEauthorblockA{MIT Lincoln Laboratory\\
Lexington, MA\\
Email: Alexia.Schulz@ll.mit.edu}}


%


\maketitle

\begin{abstract}
Spear phishing is a widespread concern in the modern network security landscape, but there are few metrics that measure the extent to which reconnaissance is performed on phishing targets.  Spear phishing emails closely match the expectations of the recipient, based on details of their experiences and interests, making them a popular propagation vector for harmful malware.  In this work we use Natural Language Processing techniques to investigate a specific real-world phishing campaign and quantify attributes that indicate a targeted spear phishing attack.  Our phishing campaign data sample comprises 596 emails -- all containing a web bug and a Curriculum Vitae (CV) PDF attachment --  sent to our institution by a foreign IP space.  The campaign was found to exclusively target specific demographics within our institution.  Performing a semantic similarity analysis between the senders' CV attachments and the recipients' LinkedIn profiles, we conclude with high statistical certainty ($p < 10^{-4}$) that the attachments contain targeted rather than randomly selected material.  Latent Semantic Analysis further demonstrates that individuals who were a primary focus of the campaign received CVs that are highly topically clustered.  These findings differentiate this campaign from one that leverages random spam.  
\end{abstract}


%
\IEEEpeerreviewmaketitle


\makeatletter{\renewcommand*{\@makefnmark}{}
\footnotetext{This paper has been accepted for publication by the IEEE Conference on Communications and Network Security in September 2015 at Florence, Italy.  Copyright may be transferred without notice, after which this version may no longer be accessible.

}
\makeatother}

Spear phishing has grown to be the predominant vector used to compromise an organization \cite{grow2011special}.  Federally Funded Research and Development Centers (FFRDCs) and University Affiliated 
Research Centers (UARCs) have seen an uptick in a specific form of phishing attempt in which r\'esum\'es or Curriculum Vitae (CVs) are sent directly to researchers by supposed job candidates.  Over the last few years 
MIT Lincoln Laboratory has become interested in one such campaign originating from a foreign IP space.  On the surface, these messages are job applications from seemingly distinguished researchers, and 
each email contains a cover letter and CV.  While it is certainly unusual that our institution - which hires exclusively US citizens - would receive applications en masse from a foreign source, the main 
point of concern in these emails is a 1-pixel image containing a malicious link.  This web bug suggests the individual or group behind these messages 
is acting in a purposefully adversarial manner.

If these emails are random spam - that is, if messages are sent randomly to targets without forethought - the situation poses little threat.  However, the adversary may be directing specific CVs to 
specific recipients in order to maximize the probability that the target will respond to the message.  If our organization is the target of such a spear phishing 
campaign, it implies the adversary is conducting reconnaissance on our colleagues in order to determine their research backgrounds or interests.  Such foreign interest in the employees of an FFRDC constitutes a security threat and merits further investigation.

According to a recent 
security report published by Symantec Corporation, the number of observed spear phishing campaigns doubled between 2012 and 2014, while the average 
number of spear phishing emails detected per day fell sharply in the same time range \cite{symantec}.  This seemingly contradictory phenomenon 
has dangerous implications:  Adversaries are possibly abandoning high-volume spear phishing in favor of more focused attacks with a heavy emphasis on detailed 
reconnaissance.  Given that most organizations' main line of defense against spear phishing attempts is training their employees to 
recognize and avoid the attacks \cite{kumaraguru2008lessons, song2014study}, an increase in the number of well-researched and convincing spear phishing 
emails would pose severe security risks in the near future.  It is therefore vital that we develop a means of characterizing the extent to which adversaries 
are profiling their phishing targets.

We seek to characterize the phishing threat in this case-study using the techniques of Natural Language Processing (NLP).  NLP provides powerful tools which are often used to compare documents or classify them into topical groups based on 
the terms in their respective vocabularies.  The adversaries provided us with a set of documents describing the research interests of their phishing identities.  If we obtain similar documents describing 
the targets, NLP allows us to visualize and quantify the similarities between adversaries and targets to determine if the attacks are more aligned with random spam or spear phishing.

The unique contribution of our work is that we provide a measure of the adversary's reconnaissance efforts and capabilities, and we use Latent Semantic 
Analysis (LSA) to gain insight into the adversary's spear phishing plan of attack.  We present NLP-based tools which can compare phishing agents to the recipients they have targeted.  We demonstrate the efficacy of our techniques using our case-study, which we are able to characterize to very high degrees of statistical certainty.  We outline strategies to improve our current methods, but maintain that these tools could identify 
adversaries' intent in similar phishing scenarios.

 

\section{Related Work}
Many researchers have tried to understand the connection between spear phishing adversaries and their targets of interest.  Van Nguyen of the Australian Government Department of Defense provided a literature survey outlining the many 
machine-learning techniques that could be used to identify and characterize an adversary \cite{nguyen2013attribution}.  Those techniques include the Latent Semantic Analysis (LSA) applied in our own 
effort.  While this reference is an exceptional summary of the tools and approaches best suited for characterization, 
it does not provide examples of the techniques in practice.

It is commonly understood that social media sites such as Facebook, Twitter, and LinkedIn can provide an adversary a wealth of information on a target's 
work interests and expertise.  As such, many researchers have examined whether there are any connections between a target's online presence and the 
phishing emails he or she received.  Dewan et al. \cite{dewan2014analyzing} studied  a set of over 4,000 targeted and 9,000 non-targeted emails and concluded that the information in the targets' LinkedIn profiles did not indicate whether the emails they received were spear phishing attempts.  However, 
Dewan et al. only analyzed the word and character counts of the LinkedIn profiles, ignoring the semantic information they provide.  In our work, we focus 
on a semantic analysis of the targets' LinkedIn profiles in order to better understand how their topical information matches that of the phishing emails.

Similar topical analyses were performed by Le Blond et al. \cite{le2014look} and DeBarr et al. \cite{debarr2013phishing}.  
When examining nearly 1,500 emails involved in a politically sensitive attack, Le Blond et al. compared the roles and identies of the phishers to those 
of the targets.  The targets' characteristics were pulled from various sources of social media data.  
This work succeeded in identifying several correlations between the adversaries and targets, but the 
characterizations of both sides were performed by manually examining all available text data 
and assigning individuals to pre-defined groups.  Our analysis, in contrast, compares the targets and adversaries automatically, which is necessary if our 
techniques are to be of any use to organizations facing incredible volumes of spear phishing attempts.  In their work, DeBarr et al. successfully applied 
NLP methods similar to our own to identify spear phishing emails.  However, their comparison was only between their phishing emails of interest and a 
previously compiled training set of known spear phishing emails.  This provides no measure of how well-matched the phishing emails were to their 
recipients.  By comparing the phishing emails to the targets' social media profiles, we are able to determine not only the presence of a spear phishing threat, 
but the degree of reconnaissance performed by the adversaries.

To the best of the authors' knowledge, our work in this paper is the first to quantitatively characterize how much effort adversaries devoted to 
researching their potential victims.  We use well-established spear phishing analysis techniques, such as LSA and characterizing 
targets via their social media profiles.  While our data set is not as extensive as many other analyses in this field, we provide a crucial first step towards 
measuring spear phishers' interest in an organization and its employees, which will certainly be of use to companies and institutions facing today's attacks 
with increased focus on in-depth reconnaissance \cite{symantec}.

\section{Data Samples}
There are three data corpora used in this analysis: an adversary dataset comprised of CV attachments to the phishing emails; a target dataset comprised of LinkedIn \cite{linkedin} profile 
information harvested on the intended victims at MIT Lincoln Laboratory; and a benchmark dataset downloaded from http://indeed.com \cite{indeed}, used to validate the
interpretation of our results.  

\subsection{Data on Adversaries}

Our sample of email attacks comprises 647 messages delivered between August 2013 and July 2014.  These attacks were sent from foreign IP address space, and each message 
was sent from one of 60 unique email accounts.  A total of 274 individuals in our organization were contacted.  In our sample, 596 emails connected a unique phisher/recipient pair and 51 were 
repeated messages.  This specific adversarial group was sending attacks before August 2013 and is still active at the time of this writing.  These messages were initially brought to our attention due to a 
suspicious web bug included in many of the emails.

Each phishing email is disguised as a job application sent by a foreign research scientist.  The salutations of the emails explicitly address the phishing target 
by name, but there is no further mention of the target or Lincoln Laboratory in the body text.  While the messages are often written in broken English, each individual seems to have a unique writing style and cites 
specific research experiences.

Attached to each email is a CV (formatted as a PDF) outlining the educational and research experiences of the supposed applicants.  
Though two of the phishing identities' CVs were corrupted and unreadable, the remaining 58 all had unique content, formatting styles, and academic fields of focus.  We extracted the plain 
text of the documents using the Apache TIKA \cite{Tika} 
parsing software.  We cross-referenced the citations listed in these CVs against online databases and verified 
that the papers were actually published by the specified authors.  Furthermore, the research experiences discussed in the CVs match those discussed within the body of the email.  Although the email address 
from which the attack was sent does not match the address listed in any of the CVs, some attempt is made to make the email look legitimate; the email handle is usually some version of the supposed job 
candidate's name.  Given the apparent legitimacy of the candidates themselves, we infer that an adversary is harvesting CVs and cover letters from research experts without their knowledge, adding easily 
faked salutations addressed to specific individuals, and sending phishing attacks from falsified email addresses.

The data used in our analysis are simply the words used within these CVs.  If some adversarial party has harvested a large, varied corpus of professional CVs and cover letters, they could feasibly attack a phishing 
target under the guise of an individual who shares the target's research interests and experiences.  The adversary and target would thus be identified as similar if their respective research profiles contain many of the same words or phrases.  We convert the CVs (and the target profiles discussed in the following section) into ``bag-of-words" models, wherein a document is characterized only by the 
presence or absence of terms from a vocabulary, not by the organization of those terms.

\subsection{Data on Targets}
\begin{figure}[b!]
\begin{center}
\includegraphics[trim=0cm 0.5cm 1cm 0.5cm, clip=true, width=0.48\textwidth, angle=0]{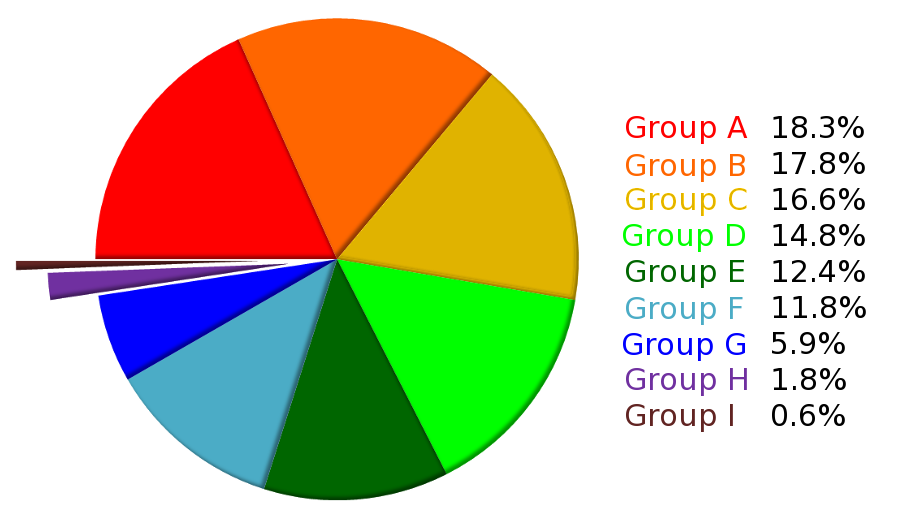}
\caption{\sl The distribution of phishing emails received by different groups at MIT Lincoln Laboratory.  Percentages are not adjusted to account for the size of each group.  Group I is small, but groups G and H 
are underrepresented in the campaign.\label{fig:pie}}
\end{center}
\end{figure}
\begin{figure}[t!]
\begin{center}
\includegraphics[trim=5cm 4cm 4cm 3cm, clip=true, width=0.5\textwidth, angle=0]{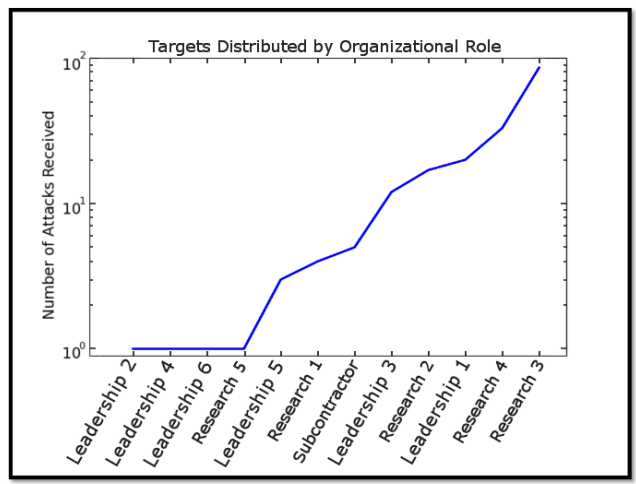}
\caption{\sl The distribution of phishing emails received by employees sorted by job title.  Titles are divided into tiers, with level 1 for employees with the least experience and responsibility, and higher levels for 
employees with more experience and responsibility.  Notably, all recipients held research or leadership positions.\label{fig:jobs}}
\end{center}
\end{figure}

To determine if the adversaries were actively researching their targets, we first examined the demographics of the phishing recipients.  MIT Lincoln Laboratory is divided into 9 distinct groups, each of 
which has its own field of expertise.  Figure \ref{fig:pie} illustrates that most groups were targeted roughly evenly by the adversary (group names have been anonymized for security reasons).  
Group I received only 0.6\% of the emails because it only contains a few members.  However, Groups G and H are disproportionately under-represented for reasons we have not yet uncovered.  Figure 
\ref{fig:jobs} illustrates how the targets are distributed by job role and hierarchical position (``Research 1" has fewer responsibilities than ``Research 2", 
and so on).  
We see that all of the targets worked as either technical research staff or as group leaders.  None of the targets were employed as secretaries, 
information technology professionals, or members of the security team.  This is especially unusual given the recent findings of the Symantec Corporation's  
Internet Security Threat Report: 39\% of employees who received a spear phishing email in 2014 were either interns 
or support staff \cite{symantec}.  Either the adversary's source of Lincoln emails contained exclusively technical personnel, or the adversary performed 
sufficient background research on their targets to determine whether or not they played a research role within the organization.

Our sample of phishing CVs provides us with textual information highlighting the interests and experiences of a group of individuals.  If the research background of the bait
matches that of the targets, we conclude that this foreign group is performing reconnaissance on members of our organization, and their attacks constitute a spear phishing campaign.  
To characterize the threat, we require a text-based research profile on each of the targets of this campaign, as well as some metric for comparing such a profile against a phishing CV and quantifying their 
similarity.

Because we did not have access to a CV for each of the 274 attack recipients, we built our sample of research profiles using publicly available information on the internet.  This 
method is not without downfalls, as searching for and gathering this information was a time-consuming manual process.  The clear advantage of this approach, however, is that it prohibits us from using any 
information which the adversaries cannot access, so our profiles are similar to what the adversaries themselves may have gathered.  While we were able to identify several online sources of research 
summaries (such as personal websites and employee web pages), it was vital that the profiles in our sample be uniformly formatted.  Otherwise, our methods prioritize similarity of data source 
over similarity of research interests, an artifact that arises because the vocabularies used by different sources all contain uniquely identifiable traits.  We determined that the research profiles 
provided by the career networking website LinkedIn \cite{linkedin} provided the most detailed information on the largest number of targets.  Much like a CV, Linked-In profiles 
often list an individual's educational history, work 
experience, skills, and sometimes publications.  After removing those which were too sparse to analyze, our final sample of research profiles contained the copied text of the LinkedIn pages
for 100 of the 274 
phishing targets.

\subsection{Benchmark Dataset}

It is critical that we benchmark the NLP characterization 
techniques used in our analysis against a data set with known topical relationships.  To perform benchmark analyses, we obtained a marked sample of CV data from the r\'esum\'e database of http://indeed.com 
\cite{indeed}.  These CVs were divided into three career categories: Postdoctoral Researchers (from any field), Managers, and Software Engineers.  We repeat every analytic test on this benchmark dataset to 
help interpret the significance of our findings about the phishing campaign.

\section{Methods and Analysis}
In this section we investigate the phishing threat posed by this campaign using well-known NLP techniques.  We have relied heavily on the
tools provided in the scikit-learn Python package \cite{scikit-learn} for our analysis. 
\subsection{n-gram Model}
We need a method of quantifying the similarity between two documents to determine whether the adversaries are selectively matching their CVs to the research backgrounds of their targets.  One common 
method of comparing documents in Natural Language Processing is known as an \emph{n-gram model}.  An n-gram is defined as a sequence of $n$ consecutive words within a document.  If a corpus of 
documents contains $M$ unique n-grams, then an n-gram model would represent each individual document as a vector of length $M$.  Each element of the vector represents a different n-gram, and the value 
of the element is a weight which represents how strongly that n-gram characterizes the document.

\begin{figure}[t!] 
\begin{center}
\includegraphics[trim=0cm 0cm 0cm 0cm, clip=true, width=0.5\textwidth, angle=0]{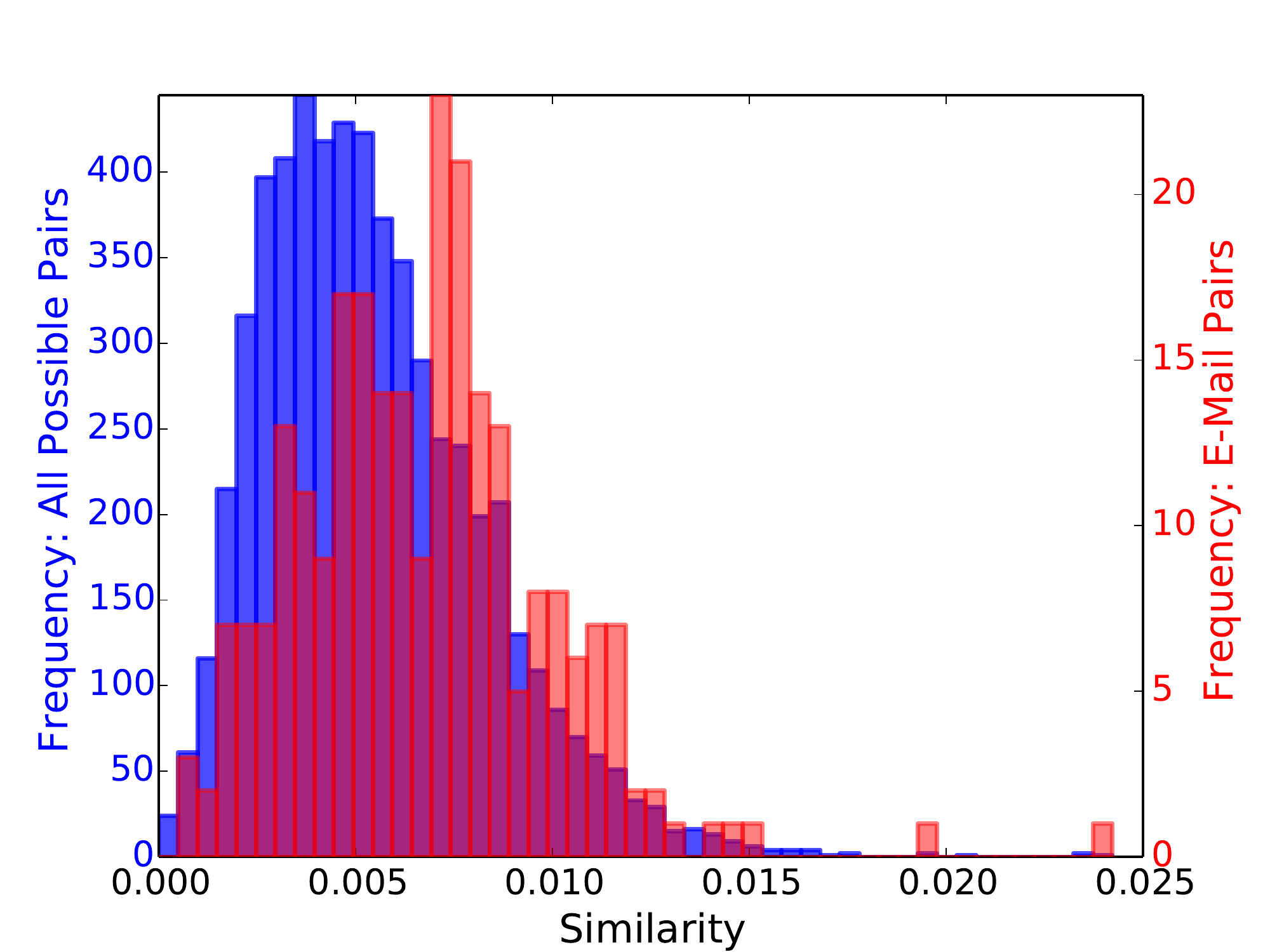}
\caption{\sl Similarity distributions for all possible phisher/target pairs (blue) and all observed pairs (translucent red).  
The median and standard error of the mean are $(5.03\pm 0.04)\times 10^{-3}$ for the blue 
histogram and $(6.65\pm 0.20)\times 10^{-3}$ for the red, indicating that the email pairs share significantly more similarities than expected from a random spamming attack. \label{fig:twohist}}
\end{center}
\end{figure}
For our analysis, we limit the value of $n$ for our n-grams to 1, 2, or 3 (unigrams, bigrams, and trigrams) and omit a pre-defined list of extremely common terms known as \emph{stop words}.  If our document 
was the simple sentence ``The quick brown fox jumped over the lazy dog," we would identify the words ``the" and ``over" as stop words, and our list of bigrams would be ``quick brown," ``brown fox," ``fox 
jumped," ``jumped lazy," and ``lazy dog."

To quantify the weight of each n-gram, we use the \emph{term frequency-inverse document frequency} value, or \emph{tf-idf}
\cite{salton1983introduction}.  If our corpus contains $N$ documents and $M$ n-grams, the tf-idf n-gram model is 
best represented as an $N \times M$ matrix where the value of element ($i$,$k$) is
\begin{equation}
\textrm{tf-idf}(i,k) = \textrm{tf}(i,k) \times \left(1 + \ln\left(\frac{N+1}{D(k)+1}\right)\right)
\label{eqn:tfidf}
\end{equation}
where tf$(i,k)$ is the term frequency of the $k^{th}$ n-gram within the $i^{th}$ document and $D(k)$ is the number of documents within the corpus which contain the $k^{th}$ n-gram.  While some tf-idf models 
define tf$(i,k)$ as the number of times n-gram $k$ appears in document $i$, we restrict the term frequency to only boolean values.  If document $i$ contains n-gram $k$, then tf$(i,k)=1$; otherwise, tf$(i,k)$=0.  
Our boolean approach is recommended for a corpus containing relatively short documents, such as those in our corpus of CVs and research profiles \cite{salton1988term}.

The second factor in Equation~\ref{eqn:tfidf} is known as the \emph{inverse document frequency}, or idf$(k)$.  Because this factor increases as $D(k)$ decreases, n-grams which appear within relatively few 
documents in our corpus will have large tf-idf weights.  An underlying assumption of our n-gram model, therefore, is that a document is best characterized by its most uncommon words.  If a word appears in only a few documents of our corpus, we would expect that the individuals described by those documents must have some similar trait which distinguishes them from the rest of the 
population.

\subsection{Similarity Analysis}
The $N \times M$ tf-idf matrix characterizes the documents of the corpus, but we require a metric to quantify the similarity between any two individual documents.  
Two $M$-element column vectors from the matrix, $\mathbf{D_A}$ and $\mathbf{D_B}$, can be extracted to form feature vectors 
that describe documents $A$ and $B$.  The similarity between the two documents is
\begin{equation}
\textrm{sim}(\mathbf{D_A},\mathbf{D_B}) = \frac{\mathbf{D_A} \cdot \mathbf{D_B}}{\lvert\mathbf{D_A}\rvert \lvert\mathbf{D_B}\rvert}
\label{eqn:sim}
\end{equation}
If two documents are very similar, we expect they share a large number of n-grams in common, and the value of their similarity will be near 1.  Two very dissimilar documents will contain 
mostly separate sets of n-grams, and their similarity will be approximately 0.  Geometrically, the similarity between two documents is the cosine of the interior angle between their two feature vectors in the 
$M$-dimensional tf-idf space.
\begin{figure}[t!]
 \begin{center}
  \includegraphics[trim=0cm 0cm 0cm 0cm, clip=true, width=0.5\textwidth, angle=0]{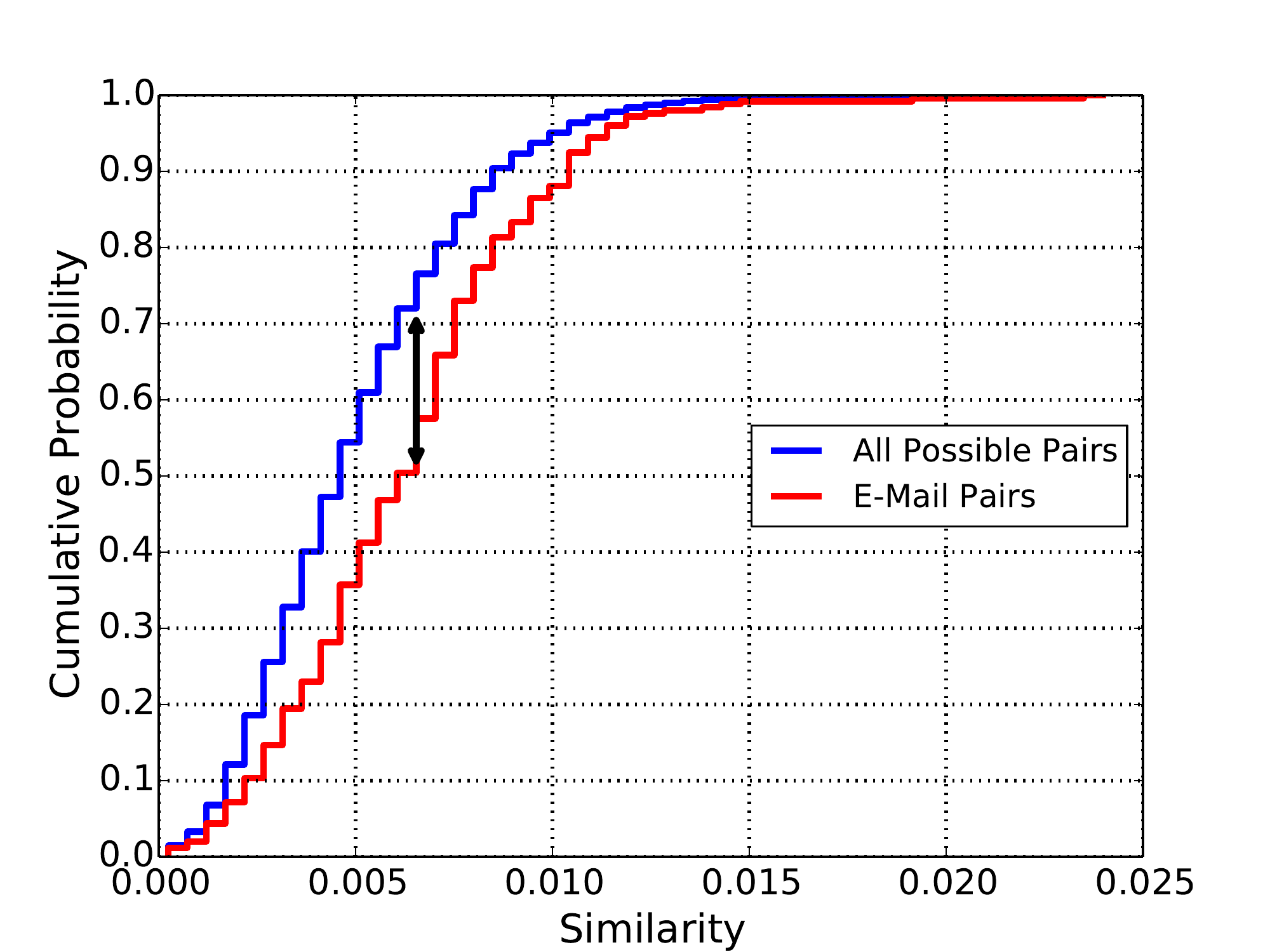}
  \caption{\sl The two-sample Kolmogorov-Smirnov test comparing the similarity distributions in Figure \ref{fig:twohist}.  Each curve is a cumulative probability distribution, and the black arrow marks their 
  supremum 
difference of 0.224, which corresponds to a probability of $<$0.01\% that the set of emails is a random sample of the set of all possible phisher/target pairs. \label{fig:kstest}}
 \end{center}
\end{figure}

To determine if the adversary is conducting a spear phishing campaign, we study certain distributions of these similarity values.  We compare two cases: the distribution of similarity values arising from the phishing 
campaign, and the distribution we would expect if CVs were sent to random recipients. We generate the latter distribution by computing the similarity between all possible phisher/recipient pairings. Since we have 58 adversary CVs and 100 target research profiles, we can calculate 5800 similarity values for this set.  The distribution of those values 
is shown in Figure~\ref{fig:twohist} as the blue histogram.  
If the adversary were haphazardly spamming our organization with emails, they would effectively select a random subset of these 5800 possible pairs, and the resulting similarity distribution would be drawn 
from the blue histogram.  In contrast, the translucent red histogram in 
Figure~\ref{fig:twohist} shows the distribution recovered when each adversary persona is paired only with the targets to whom he sent an email.  From 
the phishing identities and subset of targets with identifying documentation, we are able to calculate the similarities for 252 of these email pairs. It is immediately clear in Figure~\ref{fig:twohist} that the 
distribution of email pairs is skewed towards higher values than the distribution of all possible pairs.  The median and standard deviation of the mean for the all-pair distribution is $(5.03\pm 0.04)\times 
10^{-3}$, while those values for the email distribution are $(6.65\pm 0.20)\times 10^{-3}$.

\begin{figure}[t!]
\begin{center}
\includegraphics[trim=0cm 0cm 0cm 0cm, clip=true, width=0.5\textwidth, angle=0]{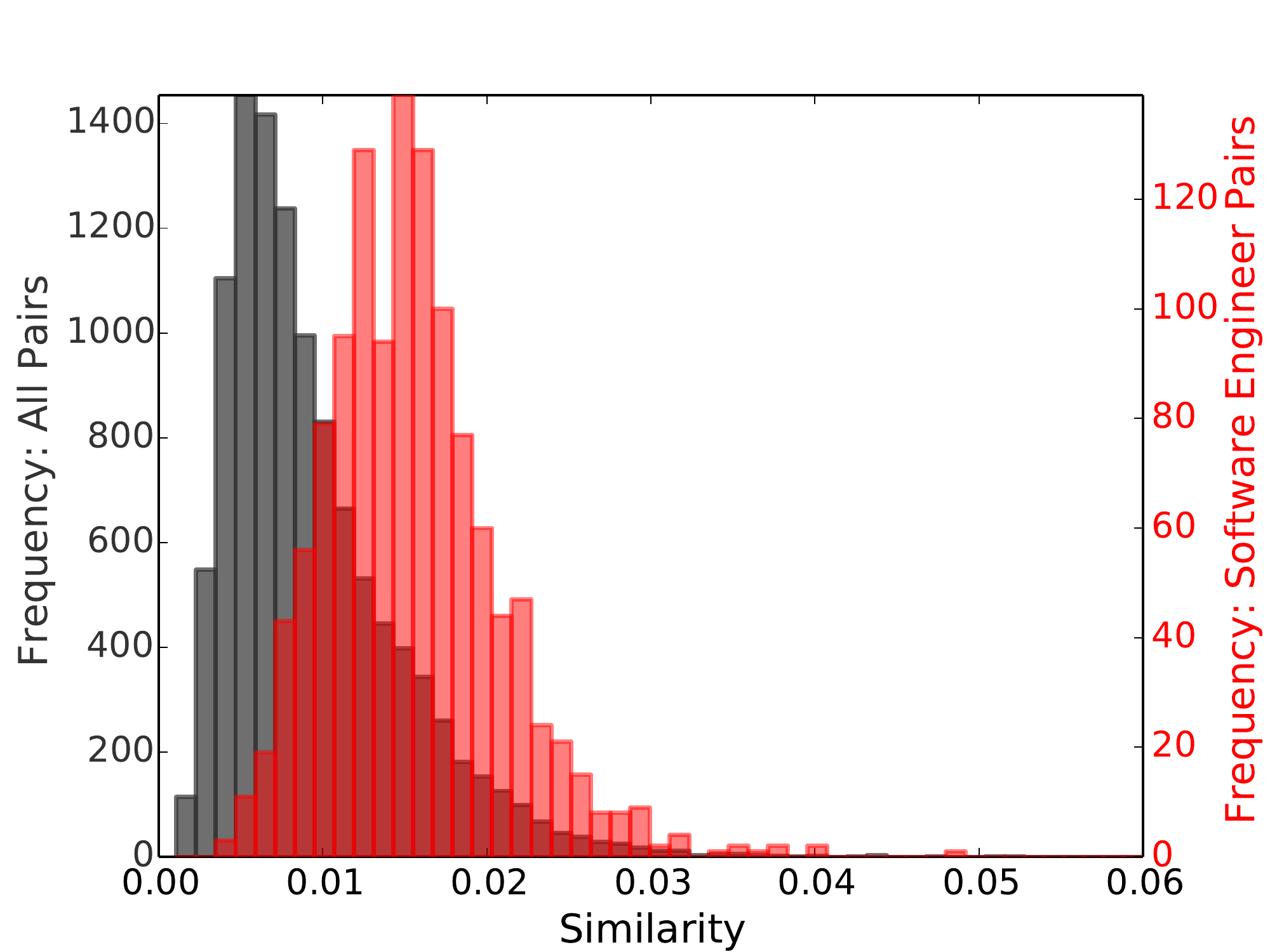}
\caption{\sl The similarity metric applied to the benchmark data sample.  The gray histogram represents the distribution from an all-to-all comparison of the documents in the 
benchmark corpus, while the translucent red 
distribution shows document comparisons among Software Engineers only.  The median and standard error of the mean are $(8.01\pm 0.05)\times 10^{-3}$ for the gray histogram and $(15.06\pm 0.20)\times 
10^{-3}$ for the red. \label{fig:benchhist}}
\end{center}
\end{figure}

To confirm that the set of email pairs is not consistent with random spamming, we compared our observed and random email similarity distributions using a two-sample Kolmogorov-Smirnov test.  The null hypothesis of 
this test is that two probability distributions were derived from the same distribution function.  The K-S statistic (the supremum difference between two cumulative probability functions, as illustrated in Figure~
\ref{fig:kstest}) for our two sets was 0.224, which allows us to reject the null hypothesis with a $p$-value less than $10^{-4}$.

Based on this result, we are certain that the adversaries did not randomly match their phishing CVs to their targets.  This implies they were instead 
profiling members of our institution and conducting a spear phishing campaign.  But how in-depth was their reconnaissance, and how effectively 
were the adversaries able to match their targets' research interests with the bait?  To gain insight into these questions, we repeated our similarity 
analysis on the indeed.com benchmark dataset.  The grey histogram in Figure~\ref{fig:benchhist} is the distribution of similarity
values arising from an all-to-all comparison of every document in the benchmark corpus, whereas the red histogram is the distribution of similarity values arising only from pairs of 
software engineers.  Just as in the case of 
the email campaign, the distribution of similarity values in the matched sample is skewed high compared to the random sample.  The median and standard deviation of the mean for the all-pair distribution is 
$(8.01\pm 0.05)\times 10^{-3}$, while those values for the email distribution are $(15.06\pm 0.20)\times 10^{-3}$. 

In both the benchmark case and the case of the phishing campaign, the median similarity of the observed pairings is significantly higher than expected for random selections.  However, it is clear that the separation between our observed email distribution and the random spamming distribution 
(Figure~\ref{fig:twohist}) is much less pronounced than the separation between the software engineer distribution and the all-career distribution 
(Figure~\ref{fig:benchhist}).  We could say that the email campaign separation is roughly $(6.65-5.03)/(0.20+0.04) = 6.75$ standard deviations, while the 
benchmark separation is $(15.06-8.01)/(0.20+0.05) = 28.2$ standard deviations.  Future application of our similarity analysis to a wide variety of 
spear phishing and spam campaigns could possibly define a formal metric from these measurements, but for now we simply state that the degree 
of reconnaissance in our case study seems to be low.

We can further characterize the adversaries' reconnaissance efforts by analyzing instances when a single CV was sent to a single target.  
For the set of all recipients who received only one email, the median similarity was $8.802 \times 10^{-3}$ with standard deviation of the mean $9.620 \times 10^{-4}$.  For the set 
of all phishers who sent only one email the median similarity value was $7.887 \times 10^{-3}$ with standard deviation $1.369 \times 10^{-3}$.  Note that some of these phishers could have sent more than 
one email, but we lack any documentation on their other recipients.  In general, these singleton instances appear to be significantly more targeted than the pairings with multiple senders or recipients.  Given that we already suspect the adversaries' profiling efforts as a whole were cursory, it is possible that their 
reconaissance efforts were focused on only a few specific individuals.

The results of this section strongly indicate that the adversary did not randomly spam Laboratory employees with emails.  Instead, we have found conclusive evidence that they have selected at least a portion 
of their phisher/target pairs on the basis of similar research or career experiences.  We deduce that the adversaries are conducting reconnaissance on employees of MIT Lincoln Laboratory, though our characterizaion of the reconnaissance suggests a limited profiling effort.

\subsection{Latent Semantic Analysis}
After discovering that our adversaries are conducting a spear phishing campaign against our organization, we wanted to learn more about how they selected their targets.  
Such information could help reveal what sort of reconnaissance this foreign group is conducting.  For instance, we could imagine a scenario in which one phishing identity is matched to a sub-group of targets, 
all of whom are semantically similar.  This could indicate the adversaries are less interested in attacking specific individuals and more interested in reaching any member of a specific research group.  
Alternatively, the adversaries could identify a single target of interest and contact that individual with many phishing identities who share his or her background.  This may imply the adversaries are conducting in-depth 
research on a few individuals in our organization, or perhaps that the phishing identities are sorted into pre-determined attack groups.
\begin{figure}[t!]
 \begin{center}
  \includegraphics[width=\columnwidth]{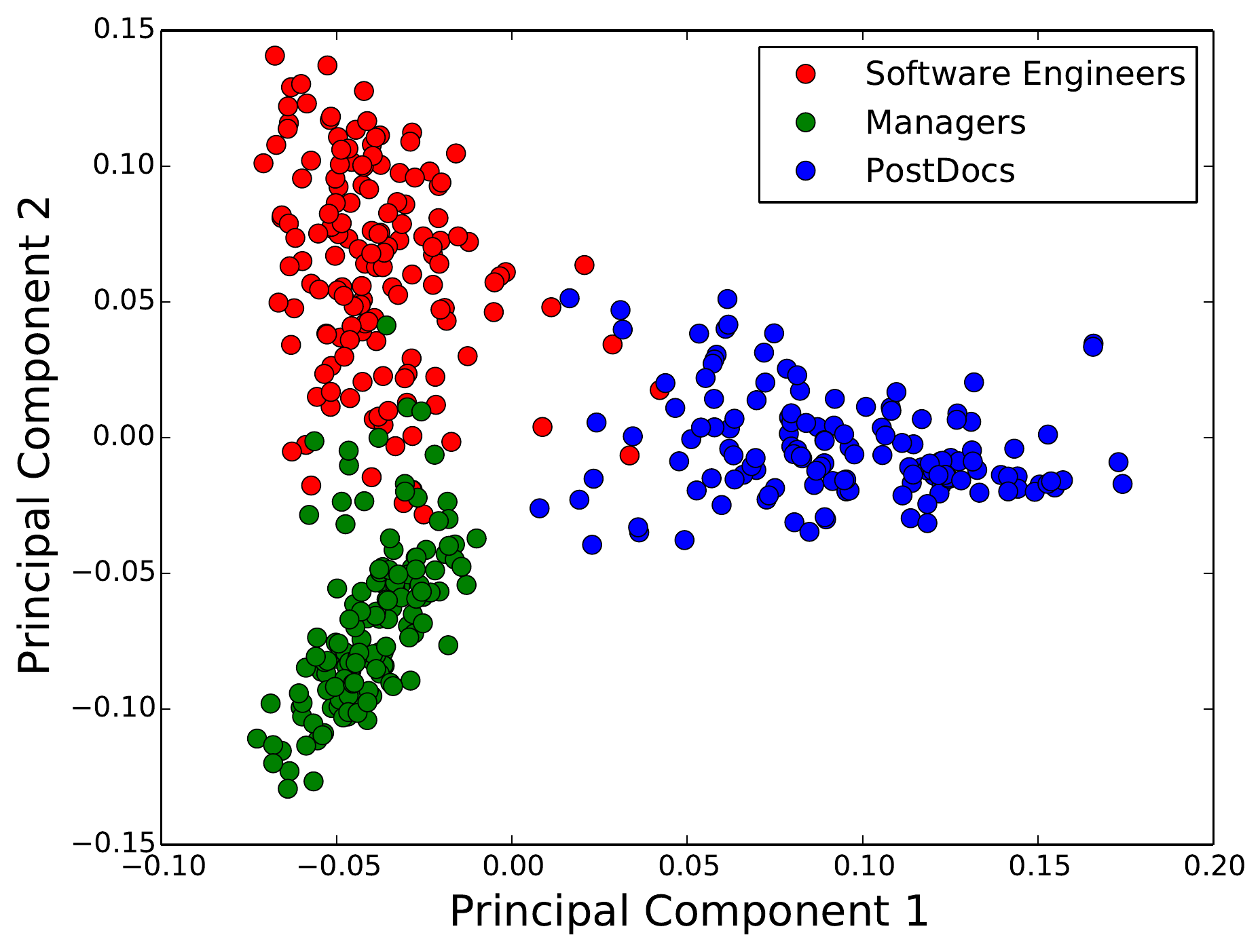}
  \caption{\sl Latent Semantic Analysis of the benchmark data set.  Plotting principle components 1 and 2 from the LSA of the indeed.com test data, we note that documents corresponding
  to Software Engineers (red), Managers (green), and Postdocs (blue) are separated into their own clusters with only slight overlap.  This illustrates that LSA can 
  reliably cluster individuals with common interests based on the content of their CVs. \label{fig:LSAdemo}}
 \end{center}
\end{figure}

Since each row of the tf-idf matrix represents a separate n-gram, and because most n-grams will not appear in any given document, our matrix has over 100,000 rows and contains mostly zeroes.  We need to 
reduce the rank of our tf-idf matrix to a dimensional space we can visualize, to determine if the phishing CVs 
and the target profiles are topically clustered.  This will indicate whether the scenarios discussed above 
apply to our problem.  We achieve this approximation using truncated Singular Value Decomposition (SVD) \cite{press2007numerical}.  
SVD is a technique by which an $m \times n$ matrix $\mathbf{M}$ is factored as $\mathbf{M} = 
\mathbf{U}\bm{\Sigma}\mathbf{V}^*$, where $\mathbf{U}$ and $\mathbf{V}$ are unitary matrices and $\bm{\Sigma}$ is an $m \times n$ rectangular diagonal matrix.  
\begin{figure*}
  \centering
  \includegraphics[trim=0cm 0cm 0cm 1.5cm, clip=true, width=0.75\textwidth]{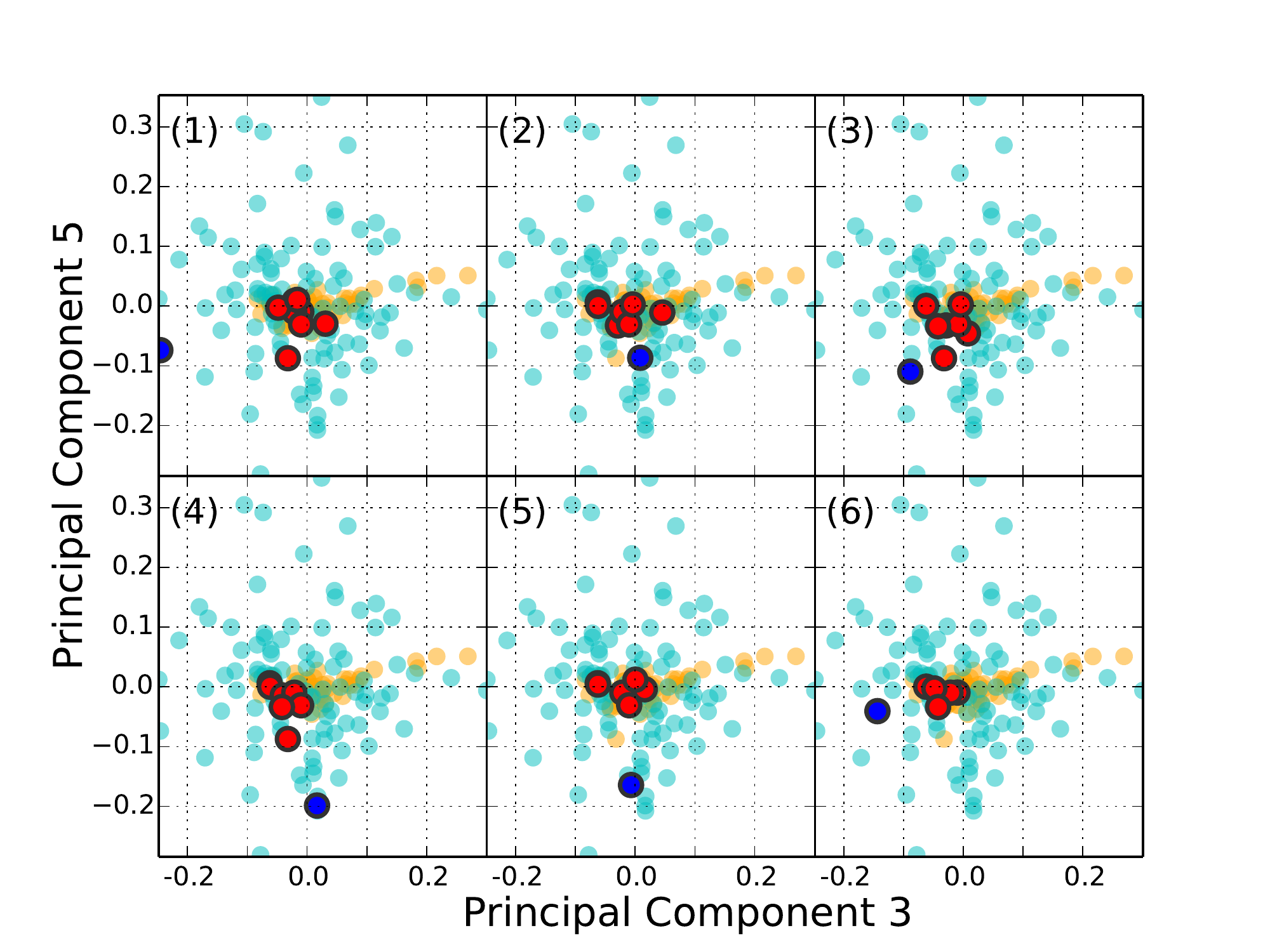}
  \caption{Principal components 3 and 5 from the Latent Semantic Analysis for groups of phishers targeting the same recipient.  Cyan or blue markers represent recipient research profiles, and orange or red markers represent phishing CVs.  Each subplot highlights one of the six targets who was most frequently attacked (blue) and the CVs they received (red).  Because the red points tend to cluster together, we assert that many similar CVs are often used to target specific individuals.}
  \label{fig:lsaphish}
\end{figure*}

\begin{figure*}
  \centering
  \includegraphics[trim=0cm 0cm 0cm 1.5cm, clip=true, width=0.75\textwidth]{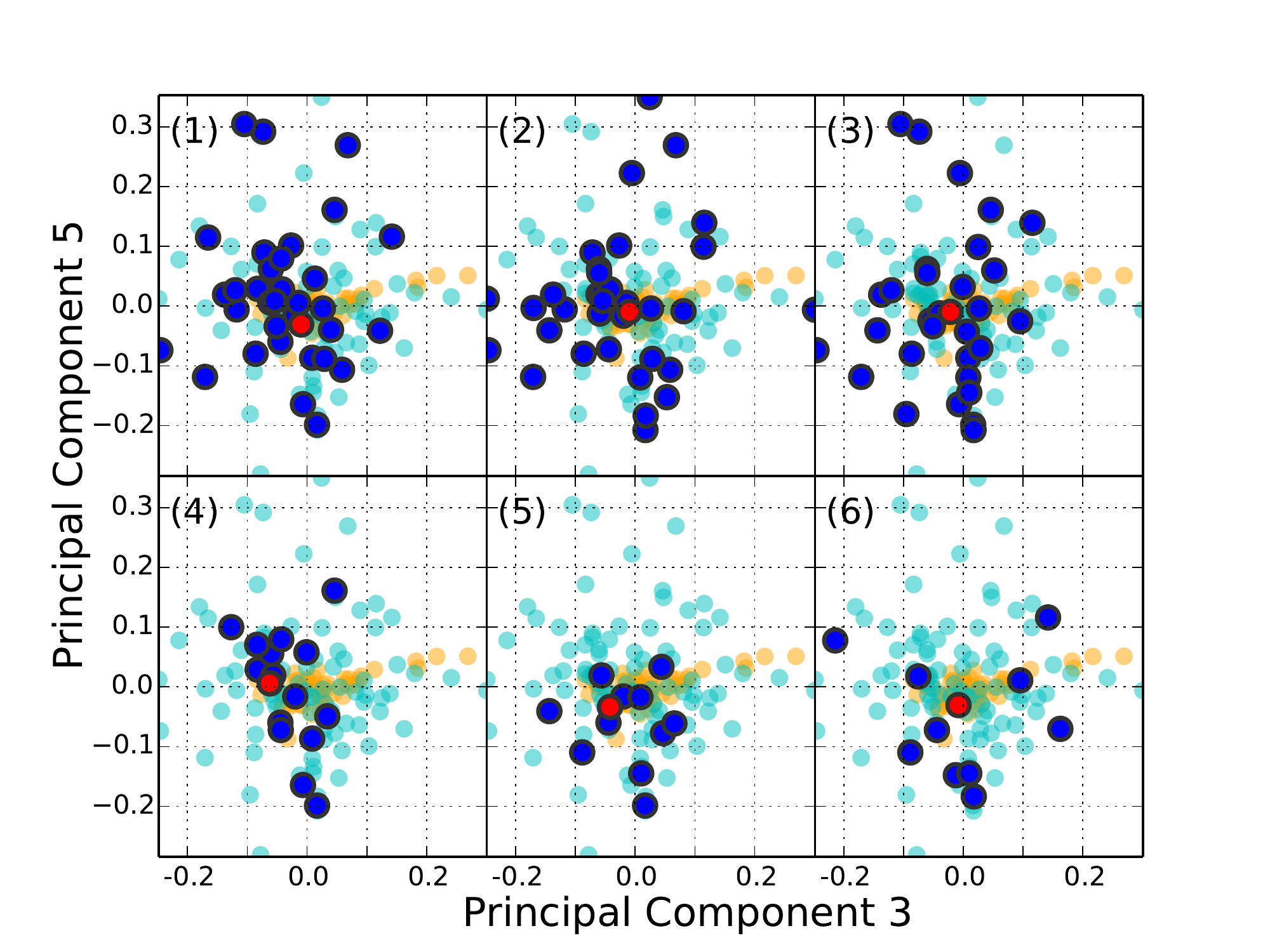}
  \caption{Principal components 3 and 5 from the Latent Semantic Analysis for groups of recipients targeted by the same phisher.  The color scheme is the same as in Figure \ref{fig:lsaphish}, but here we highlight the six phishers who sent the most attacks (red) and their respective groups of recipients (blue).  There seems to be no clustering or correlation between the blue points, implying that highly active phishers do not attack specific groups of targets.}
  \label{fig:lsarecip}
\end{figure*}


\begin{figure*}
\centering
\subfigure{\label{fig:wsub1}\includegraphics[width=\columnwidth]{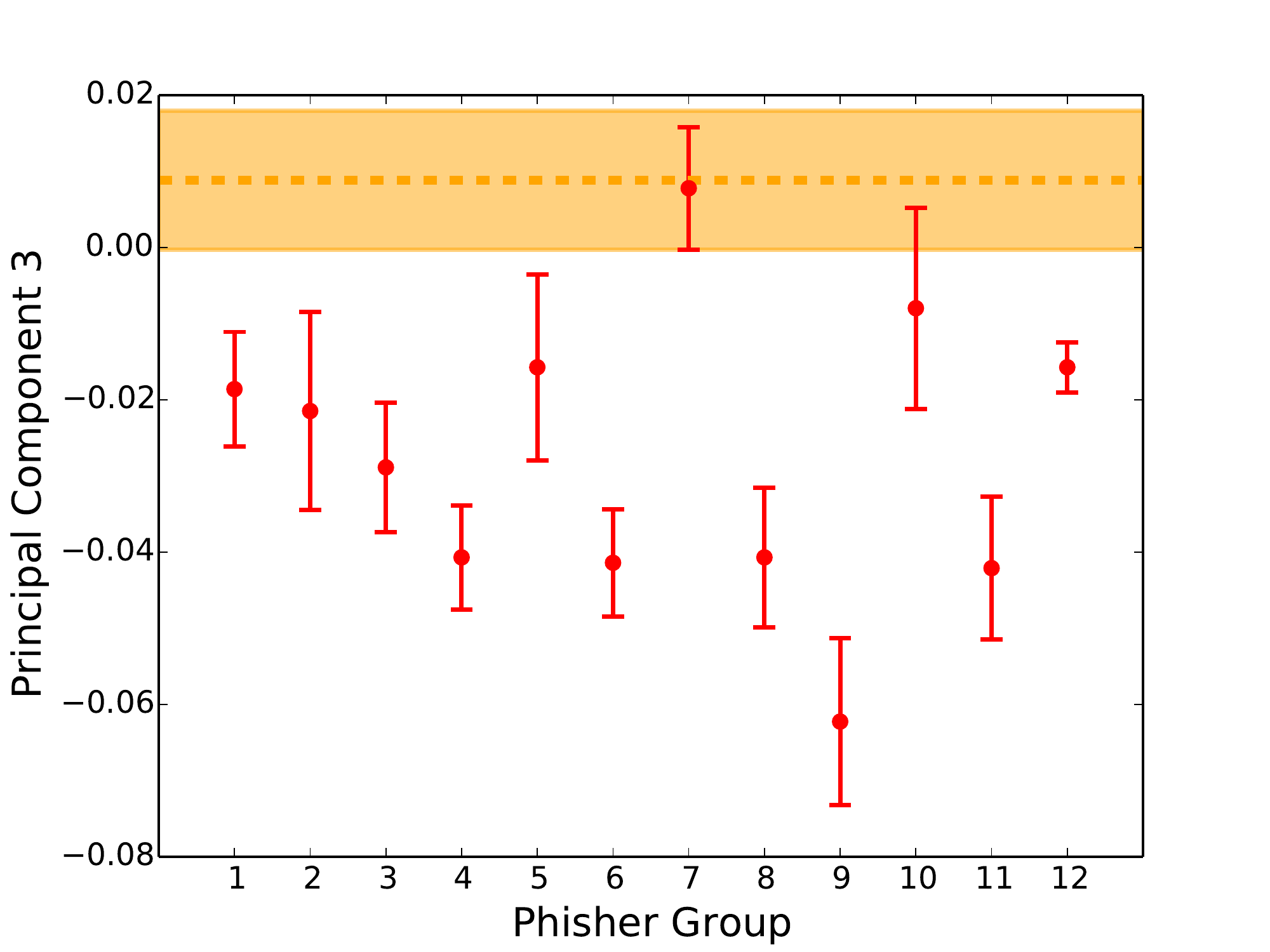}}
\subfigure{\label{fig:wsub2}\includegraphics[width=\columnwidth]{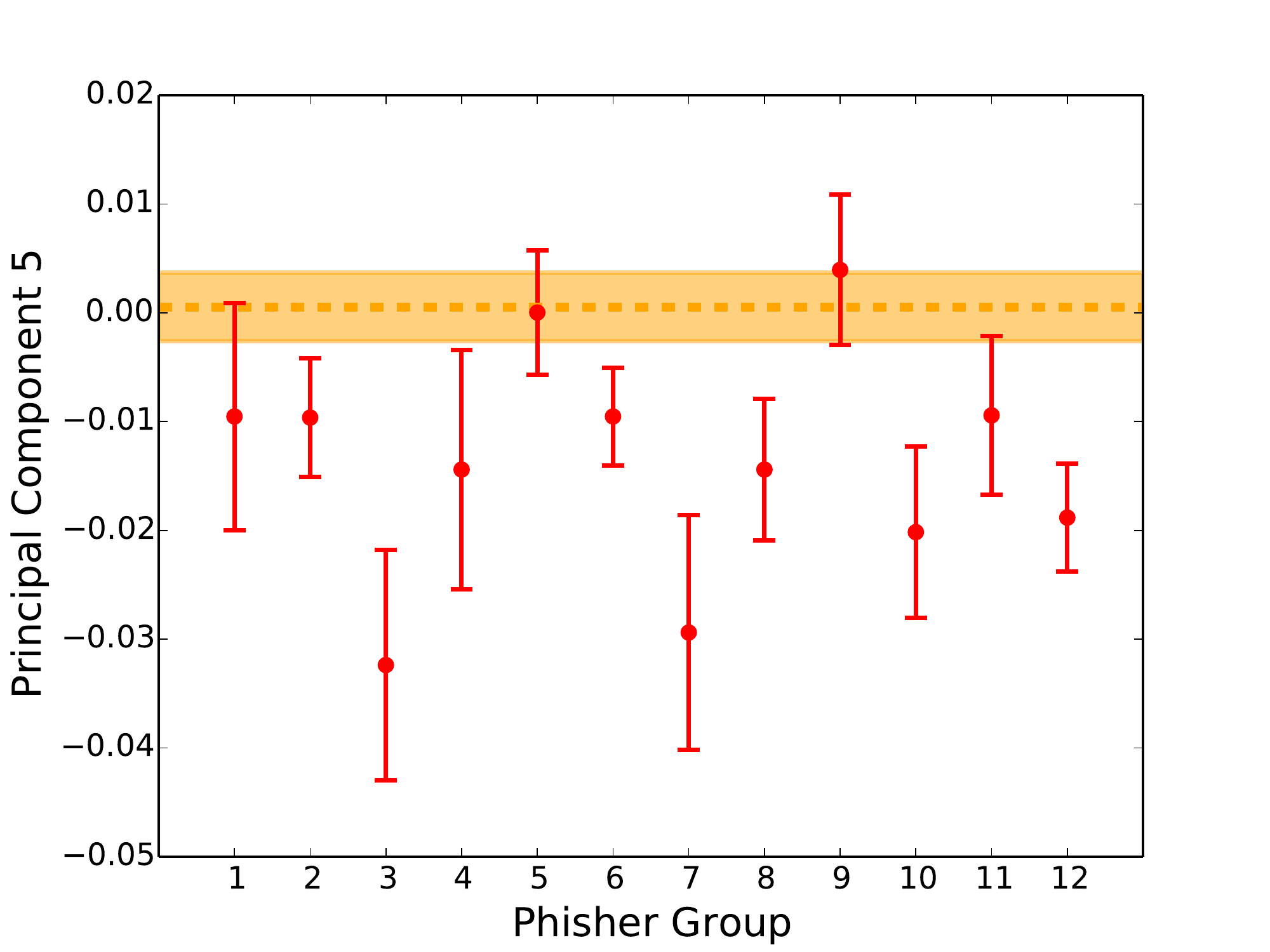}}
\caption{Search for clustering in individual Principal Components: phisher groups.  The horizontal dashed line  shows the median value of PC3 (left) or PC5 (right) for all the phishing CVs, and the shaded area represents the standard error of the mean for the same distributions.  Each red marker and error bar shows the median and standard error of the mean for the group of phishers who attacked one of the 12 most frequently targeted recipients.  The significant and systematic offset of the red data points implies that these groups of phishers are not representative of the phisher population.  Instead, they form clusters as seen in Figure \ref{fig:lsaphish}.}
\label{fig:whiskers1}
\end{figure*}

\begin{figure*}
\centering
\subfigure{\label{fig:wsub3}\includegraphics[width=\columnwidth]{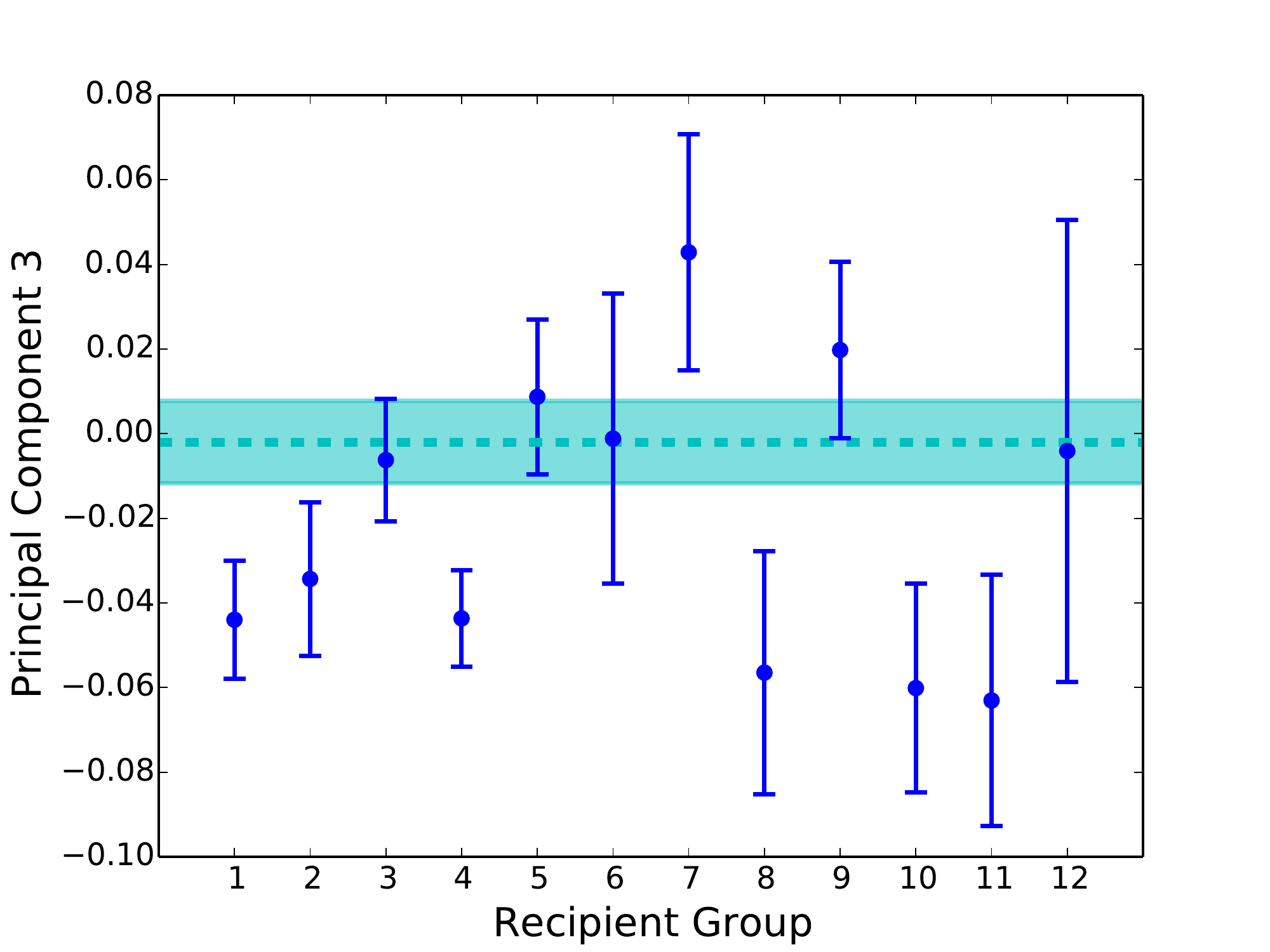}}
\subfigure{\label{fig:wsub4}\includegraphics[width=\columnwidth]{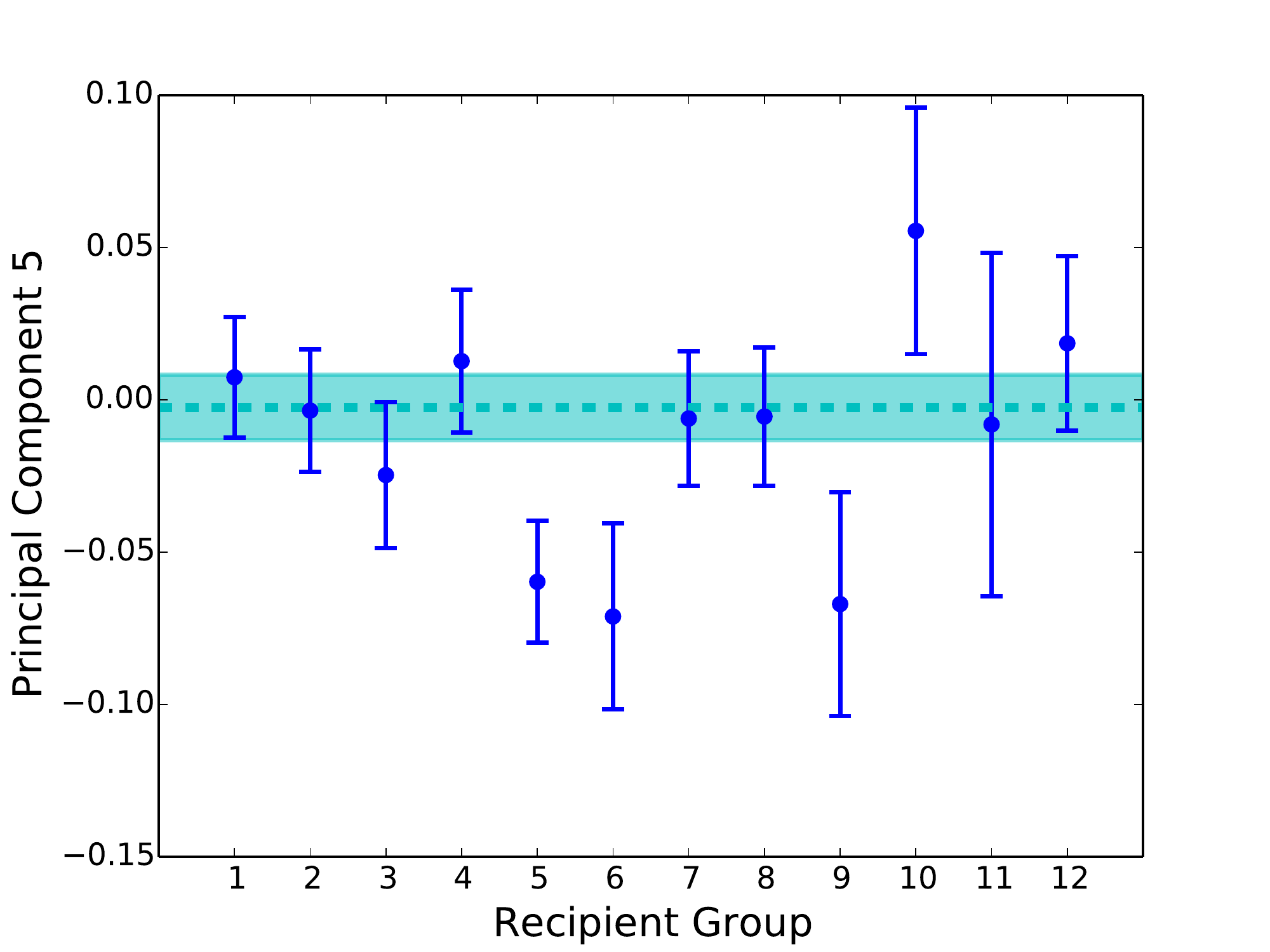}}
\caption{Search for clustering in individual Principal Components: recipient groups.  This figure is the same as Figure \ref{fig:whiskers1}, but the cyan dashed line and shaded region represent the median and standard error of the mean for all recipient research profiles, and the blue markers and error bars represent the same features for the groups of recipients attacked by the 12 most active phishers.  Unlike in Figure \ref{fig:whiskers1}, there is no systematic offset of the blue data points, and the errors are similar to or greater than the population error, so we see no evidence of clustering.}
\label{fig:whiskers2}
\end{figure*}

The diagonal values of $\bm{\Sigma}$, 
conventionally ordered from greatest to least, are non-negative real numbers known as singular values.  According the the Eckart-Young theorem 
\cite{salton1975vector}, the closest possible $k$-rank approximation to $
\mathbf{M}$ can be calculated by using only the $k$ columns of $\mathbf{U}$ and $k$ rows of $\mathbf{V}$ corresponding to the $k$ greatest singular values, yielding the matrix $\mathbf{\tilde{M}} = 
\mathbf{U_k}\bm{\Sigma_k}\mathbf{V^*_k}$.  This process is known as truncated SVD, and the application of truncated SVD upon a tf-idf matrix is known as \emph{Latent Semantic Analysis} (LSA) 
\cite{deerwester1990indexing}.

Before applying this method to analyze the phishing campaign, we tested how effectively LSA could separate a corpus of CVs into topical clusters. 
To perform this benchmark, we
used the marked sample of CV data from http://indeed.com \cite{indeed}, divided into three categories: postdocs, 
managers and software engineers.  Figure \ref{fig:LSAdemo} demonstrates that 
CVs with a known relationship are significantly clustered in at least one projection onto the singular vectors (principle components or PCs).

Returning to the analysis of the phishing campaign, after performing LSA with $k=100$ on our tf-idf matrix we obtain a matrix with 158 columns (one for each phishing CV and target research profile) and 100 
rows.  Each row is now a principal component vector which represents a linear combination of n-grams.  For instance, the first component (PC1) assigns strong negative weights to n-grams such as ``lincoln 
laboratory" and ``greater boston area" and strong positive weights to terms such as ``curriculum vitae" and the name of the phishing group's home country.  As a result, PC1 efficiently groups documents into 
adversary and target groups.  By plotting two principal components against each other, we can visualize how each of the 158 documents are related in that two-dimensional cross-section of semantic space.  If 
any given group of documents share strong similarities, they will tend to form clusters in at least some of these plots.

In Figures \ref{fig:lsaphish} and \ref{fig:lsarecip}, we plot PC5 against PC3.  We selected this subspace because it spanned a semantically interesting subset of terms.  For PC3,
strong negative weight is assigned to n-grams such as ``artificial intelligence," ``machine learning," and ``statistical," while strong positive weight is given to terms 
like ``optics," ``molecular,"
and ``quantum."  So among other things
 PC3 seems to distinguish computer science from physics.  Likewise PC5 positively weights the terms ``computer science" and ``national laboratory research" and 
negatively weights ``radar" and ``sensor."  It's important to note, however, that these are just a few example terms, and each principle component is actually a mixture of several 
different research areas. 

Each data point in Figures \ref{fig:lsaphish} and \ref{fig:lsarecip} represents a single document, with orange or red markers for phishing CVs and blue or cyan markers for target profiles.  Each subplot in Figure 
\ref{fig:lsaphish} highlights one of the six targets who received the largest number of phishing emails (blue marker) and the group of phishing CVs with which they were targeted (red markers).  Figure 
\ref{fig:lsarecip} uses the same highlighting scheme to show the 6 phishing CVs which were most frequently sent out and the targets who received them.  We see in Figure \ref{fig:lsarecip} that the targets of 
these high-volume phishers seem to be randomly scattered about this principal component space with no apparent clustering.  On the other hand, the phishing groups targeting single individuals in each subplot 
of Figure \ref{fig:lsaphish} tend to form tight clusters.  Most notably, the tail of phishing CV data points that stretches out towards positive values of PC3 -- \emph{away} from the targets -- does not include a 
single red data point in any plot.

These clustering observations are further supported by Figures \ref{fig:whiskers1} and \ref{fig:whiskers2}, where we focus on one principal component at a time.  The dashed horizontal line in each plot 
represents the median PC value of \emph{all} phishing CVs (orange) or \emph{all} target profiles (cyan).  The shaded areas around the dashed lines show the standared error of the mean.  Each data point and 
error bar shows the median and standard error of the mean for the phishing groups attacking a single target (red) or the group of recipients of a single phisher (blue).  In decreasing order from left to 
right, the data points represent the twelve largest groups attacking one recipient or the twelve largest target groups attacked by a single phisher.  For both PC3 and PC5 in Figure \ref{fig:whiskers2}, we see that 
no points are significantly removed from the population median, and their errors are similar to or larger than the population error.  These observations are in agreement with the lack of clustering seen in Figure 
\ref{fig:lsarecip}.  However, the data points in Figure \ref{fig:whiskers1} show systematic removal from the population median, often with small deviations in PC3.  The groups of red data points in Figure 
\ref{fig:lsaphish} are indeed forming clusters which are not representative samples of the population.

These findings reveal the adversaries' general plan of attack: Any single phisher may attack a large number of unrelated targets by itself, but groups of phishing CVs are often selected to make a coordinated 
attack on a target with whom they share greater-than-average similarity.  This approach could be effective if the adversaries' plan is to maximize the probability that a specific target will eventually respond to a 
message.  The clustering observed in PC3 and PC5 is not observed in all principal components, which is in agreement with Figure \ref{fig:twohist}'s demonstration that even the most similar phisher/target pairs 
are still quite different.  Nonetheless, the clustering evident in Figure \ref{fig:lsaphish} is apparently sufficient to explain the statistically significant similarity distribution separations we found in the previous section.

\section{Conclusion}
Through the application of Natural Language Processing, we successfully identified and characterized a particular spear phishing campaign against our organization.  Our data sample comprised a set of 58 Curriculum Vitae received from the adversary, and the text from Linked-In profiles belonging to 100 of the targeted individuals.  We demonstrated with a high degree of statistical certainty ($p < 10^{-4}$) that the phishing identities used to contact the targets were, on average, more similar to their respective recipients than expected if the pairings were randomly chosen.  We found that the median of the 
email pair similarities was removed from the median of all-possible-pair similarities by 6.75 standard deviations, which is relatively minor compared to 
our benchmark analysis (28.2 standard deviations), implying the adversary's reconnaissance capabilities were somewhat limited.  Finally, we applied Latent Semantic Analysis to explore the clustering of the phishing identities and their targets in one principal component subspace, finding clear evidence that individual recipients were often contacted by a sub-group of semantically similar attackers.

Our analysis has revealed that an adversarial group is performing reconnaissance on our colleagues, and we have a basic understanding of this phishing group's plan of attack.  However, our work could be improved upon in many aspects.  First, our data sample could be expanded, both by retrieving phishing CVs from earlier time periods and by obtaining research profiles for more of the targets.  Furthermore, we would like to investigate further to determine if the adversaries have a specific algorithm for selecting which targets to contact using which identities.  We may be able to achieve this by analyzing if and how the relationships between adversary and target change over time and across principal component space.  Repeating our analysis over time could also detect 
if the adversary's interest is escalating, as indicated by increasing similarity between the target profiles and the bait.

Nonetheless, our work provides the first quantitative characterization of a spear phishing adversary's reconnaissance efforts.  Our tools are highly automated, 
and none of our techniques or methods were exclusively applicable to our case study.  As long as an analyst is able to obtain text-based profiles characterizing both the phishing identities and their targets, this similarity analysis and Latent Semantic Analysis could be applied to any 
spear phishing problem to identify the adversaries' intent and capabilities.


\section*{Acknowledgment}

The authors wish to thank Kara Greenfield, Bill Campbell, Robert Hall, and Sean Cavanaugh for their sage advice and suggestions.

This work is sponsored by the Assistant Secretary of Defense for Research \& Engineering under Air Force Contract \#FA8721-05-C-0002. 
Opinions, interpretations, conclusions and recommendations are those of the authors and are not necessarily endorsed by the United States Government.




\begin{thebibliography}{10}
\providecommand{\url}[1]{#1}
\csname url@samestyle\endcsname
\providecommand{\newblock}{\relax}
\providecommand{\bibinfo}[2]{#2}
\providecommand{\BIBentrySTDinterwordspacing}{\spaceskip=0pt\relax}
\providecommand{\BIBentryALTinterwordstretchfactor}{4}
\providecommand{\BIBentryALTinterwordspacing}{\spaceskip=\fontdimen2\font plus
\BIBentryALTinterwordstretchfactor\fontdimen3\font minus
  \fontdimen4\font\relax}
\providecommand{\BIBforeignlanguage}[2]{{%
\expandafter\ifx\csname l@#1\endcsname\relax
\typeout{** WARNING: IEEEtran.bst: No hyphenation pattern has been}%
\typeout{** loaded for the language `#1'. Using the pattern for}%
\typeout{** the default language instead.}%
\else
\language=\csname l@#1\endcsname
\fi
#2}}
\providecommand{\BIBdecl}{\relax}
\BIBdecl

\bibitem{grow2011special}
B.~Grow and M.~Hosenball, ``Special report: In cyberspy vs. cyberspy, china has
  the edge,'' 2011.

\bibitem{symantec}
P.~Wood, Ed., \emph{Internet Security Threat Report 2015}, ser. Internet
  Security Threat Report (Symantec).\hskip 1em plus 0.5em minus 0.4em\relax
  Symantec Corporation.

\bibitem{kumaraguru2008lessons}
P.~Kumaraguru, S.~Sheng, A.~Acquisti, L.~F. Cranor, and J.~Hong, ``Lessons from
  a real world evaluation of anti-phishing training,'' in \emph{eCrime
  Researchers Summit, 2008}.\hskip 1em plus 0.5em minus 0.4em\relax IEEE, 2008,
  pp. 1--12.

\bibitem{song2014study}
M.~Song, J.~Seo, and K.~Lee, ``Study on the effectiveness of the security
  countermeasures against spear phishing,'' in \emph{Information Security
  Applications}.\hskip 1em plus 0.5em minus 0.4em\relax Springer, 2014, pp.
  394--404.

\bibitem{nguyen2013attribution}
V.~Nguyen, ``Attribution of spear phishing attacks: A literature survey,'' DTIC
  Document, Tech. Rep., 2013.

\bibitem{dewan2014analyzing}
P.~Dewan, A.~Kashyap, and P.~Kumaraguru, ``Analyzing social and stylometric
  features to identify spear phishing emails,'' in \emph{Electronic Crime
  Research (eCrime), 2014 APWG Symposium on}.\hskip 1em plus 0.5em minus
  0.4em\relax IEEE, 2014, pp. 1--13.

\bibitem{le2014look}
S.~Le~Blond, A.~Uritesc, C.~Gilbert, Z.~L. Chua, P.~Saxena, and E.~Kirda, ``A
  look at targeted attacks through the lense of an ngo,'' in \emph{Proceedings
  of the 23rd USENIX conference on Security Symposium}.\hskip 1em plus 0.5em
  minus 0.4em\relax USENIX Association, 2014, pp. 543--558.

\bibitem{debarr2013phishing}
D.~DeBarr, V.~Ramanathan, and H.~Wechsler, ``Phishing detection using traffic
  behavior, spectral clustering, and random forests,'' in \emph{Intelligence
  and Security Informatics (ISI), 2013 IEEE International Conference on}.\hskip
  1em plus 0.5em minus 0.4em\relax IEEE, 2013, pp. 67--72.

\bibitem{linkedin}
``Linkedin; https://www.linkedin.com.''

\bibitem{indeed}
``Indeed: one search, all jobs; http://www.indeed.com/.''

\bibitem{Tika}
``Apache tika; http://tika.apache.org/.''

\bibitem{scikit-learn}
F.~Pedregosa, G.~Varoquaux, A.~Gramfort, V.~Michel, B.~Thirion, O.~Grisel,
  M.~Blondel, P.~Prettenhofer, R.~Weiss, V.~Dubourg, J.~Vanderplas, A.~Passos,
  D.~Cournapeau, M.~Brucher, M.~Perrot, and E.~Duchesnay, ``Scikit-learn:
  Machine learning in {P}ython,'' \emph{Journal of Machine Learning Research},
  vol.~12, pp. 2825--2830, 2011.

\bibitem{salton1983introduction}
G.~Salton and M.~J. McGill, ``Introduction to modern information retrieval,''
  1983.

\bibitem{salton1988term}
G.~Salton and C.~Buckley, ``Term-weighting approaches in automatic text
  retrieval,'' \emph{Information processing \& management}, vol.~24, no.~5, pp.
  513--523, 1988.

\bibitem{press2007numerical}
W.~H. Press, \emph{Numerical recipes 3rd edition: The art of scientific
  computing}.\hskip 1em plus 0.5em minus 0.4em\relax Cambridge university
  press, 2007.

\bibitem{salton1975vector}
G.~Salton, A.~Wong, and C.-S. Yang, ``A vector space model for automatic
  indexing,'' \emph{Communications of the ACM}, vol.~18, no.~11, pp. 613--620,
  1975.

\bibitem{deerwester1990indexing}
S.~C. Deerwester, S.~T. Dumais, T.~K. Landauer, G.~W. Furnas, and R.~A.
  Harshman, ``Indexing by latent semantic analysis,'' \emph{JASIS}, vol.~41,
  no.~6, pp. 391--407, 1990.

\end{thebibliography}
%
%
%


\end{document}